\documentclass{elsart}
\usepackage{graphicx}

\begin{document}
\begin{frontmatter}

\title{A Search for Three and Four Point Correlation in HiRes Stereo Data} 
\author[utah]{R.U.~Abbasi\corauthref{cor1},}
\author[utah]{T.~Abu-Zayyad,}
\author[lanl]{J.F.~Amman,}
\author[utah]{G.C.~Archbold,}
\author[utah]{K.~Belov,}
\author[utah]{S.A.~Blake,}
\author[utah]{J.W.~Belz,}
\author[rut]{D.R.~Bergman,}
\author[utah]{G.W.~Burt,}
\author[utah]{Z.~Cao,}
\author[utah]{W.~Deng,}
\author[utah]{Y.~Fedorova,}
\author[utah]{J.~Findlay,}
\author[col]{C.B.~Finley,}
\author[utah]{R.C.~Gray,}
\author[utah]{W.F.~Hanlon,}
\author[lanl]{C.M.~Hoffman,}
\author[lanl]{M.H.~Holzscheiter,}
\author[rut]{G.A.~Hughes,}
\author[lanl]{P.~H\"{u}ntemeyer,}
\author[utah]{B.F~Jones,}
\author[utah]{C.C.H.~Jui,}
\author[utah]{K.~Kim,}
\author[umt]{M.A.~Kirn,}
\author[utah]{E.C.~Loh,}
\author[utah]{M.M.~Maestas,}
\author[japan]{N.~Manago,}
\author[lanl]{L.J.~Marek,}
\author[utah]{K.~Martens,}
\author[unm]{J.A.J.~Matthews,}
\author[utah]{J.N.~Matthews,}
\author[utah]{S.A.~Moore,}
\author[col]{A.~O'Neill,}
\author[lanl]{C.A.~Painter,}
\author[rut]{L.~Perera,}
\author[utah]{K.~Reil,}
\author[utah]{R.~Riehle,}
\author[unm]{M.~Roberts,}
\author[utah]{D.~Rodriguez,}
\author[japan]{M.~Sasaki,}
\author[rut]{S.~Schnetzer,}
\author[rut]{L.M.~Scott,}
\author[lanl]{G.~Sinnis,}
\author[utah]{J.D.~Smith,}
\author[utah]{P.~Sokolsky,}
\author[col]{C.~Song,}
\author[utah]{R.W.~Springer,}
\author[utah]{B.T.~Stokes,}
\author[utah]{J.R.~Thomas,}
\author[utah]{S.B.~Thomas,}
\author[rut]{G.B.~Thomson,}
\author[lanl]{D.~Tupa,}
\author[utah]{L.R.~Wiencke,}
\author[rut]{A.~Zech,}
\author[col]{X.~Zhang,}

\address[utah]{University of Utah,
Department of Physics and High Energy Astrophysics Institute,
Salt Lake City, UT, USA}
\address[lanl]{Los Alamos National Laboratory, Los Alamos, NM, USA}
\address[umt]{Montana State University, Department of Physics, Bozeman, Montana, USA}
\address[rut]{The State University of New Jersey, Department of Physics and Astronomy, Piscataway, New Jersey, USA}
\address[col]{Columbia University, Department of Physics and Nevis Laboratory, New York, New York, USA}
\address[unm]{University of New Mexico, Department of Physics and Astronomy, Albuquerque, New Mexico, USA}
\address[japan]{University of Tokyo, Institute for Cosmic Ray Research, Kashiwa, Japan}
\corauth[cor1]{
Corresponding~author.\ {\it E-mail~address}:~rasha@cosmic.utah.edu
}

\newpage

\begin{abstract}
In this paper we investigate the possible existence of multi-point correlation in the arrival direction of the UHECR events detected by the High Resolution Fly's Eye (HiRes) stereo detector. Multi-point correlations could result from the deflection of UHECRs by galactic and intergalactic magnetic fields, and the subsequent dispersion of arrival directions from point like sources. The search is performed by calculating the solid angle subtended by the polygon between triplets and quadruplets of events in the HiRes data. The resulting distribution of solid angles is then compared to the cumulative distributions from multiple simulated isotropic data sets to estimate the significance of any excess. We also looked for potential correlation of the small solid angle triplets found in the data with the locations of the BL Lacertae (BL Lac) objects. Neither statistically significant clustering nor significant correlations with BL Lac objects were found in these studies.  
\end{abstract}


\end{frontmatter}
 
\section{Introduction}

While there have been reports of possible correlation, the sources of UHECRs are currently unknown (\cite{chad},~\cite{agasa},~\cite{auger}). One of the challenges in answering this question lies within our limited knowledge of the galactic, inter- and extra-galactic magnetic fields. Deflections of UHECR particles in the magnetic fields complicates the correlation that should otherwise exist between the arrival direction of these particles and the source position in the sky. However, moderate deflections of particles from the same source through the magnetic field should result in clustering in their arrival direction. 

Clustering in the arrival direction of UHECRs using the HiRes stereo data set has been investigated in a previous study searching for events in the sky with an arrival direction lying on a great circle~\cite{hough}.  In this paper, however, the clustering search uses the solid angle subtended by polygons in the data. This method does not look for a particular structure (e.g. an arc). The search using this method could involve a coherent or an incoherent clustering of events. 

In addition to looking for events clustering, a search for correlation with BL Lac objects is performed.  BL Lacs are blazars, a type of Active Glactic Nuclei (AGN) with its jets aligned along our line of sight. Blazars are an established source of TeV Gamma rays~\cite{ch6-3} and they are considered highly motivated sources for UHECRs.

Recent studies (\cite{ch6-4},~\cite{ch6-5},~\cite{ch6-6},~\cite{ch6-7}) have reported correlations between the AGASA, Yakutsk, and HiRes cosmic ray data sets and a subset of BL Lac objects. In addition, a study in~\cite{chad} verified the correlation between the HiRes stereo data set and a subset of BL Lac objects (BL Lacs with magnitudes $< 18$, including the Highly Polarized [HP] BL Lacs). However, and as mentioned in~\cite{chad} the significance of the observation is difficult to estimate due to the apparent $\it{a~posteriori}$ nature of the initial search. These reports have yet to be confirmed with a clean, independent data set. 

In this paper we apply alternative methods to examine this HiRes stereo data set for three- and four-point correlations as well as possible correlation between clusters found in the data and BL Lac objects. The data set used in this paper is the High Resolution Fly's Eye (HiRes) stereo data set collected between 12/1999 and 1/2004 and consists of 271 events with energy $> 10^{19}$~eV.  This is the same data set used in (\cite{chad},~\cite{data}). Particles with lower energies would likely suffer deflections large enough to destroy the correlation between the arrival direction and source position. The angular resolution of the HiRes stereo detector for this data set is $\sim 0.6^{\circ}$.  More details about this data are given in~\cite{data}.  

\section{Search for Three-point correlation}

In order to identify statistically significant clustering, we look for excess in the real data over what is found in simulated isotropic background sets. The simulation data sets used for this analysis are based on a Monte Carlo that accounts for full geometrical and exposure acceptance of the detector. A detailed description of this MC can be found in~\cite{data}.

We examined every possible combination of three events in the data set. However, because we are looking for clustering around a point source, we concentrate on those triplets subtending small solid angles. Small solid angles formation could imply the presence of an arc formation searched for in~\cite{hough} and/or clustering in the arrival direction of detected UHECRs. To this end, we restrict our search to have angular separation between any pair of events within a triplet of less than $10^{\circ}$. The solid angle $\Omega$, is calculated using Girard's theorem (\cite{ch8},~\cite{ch8-1}).

\begin{figure}[!h]
\begin{center}
{\includegraphics[width=12.0cm]{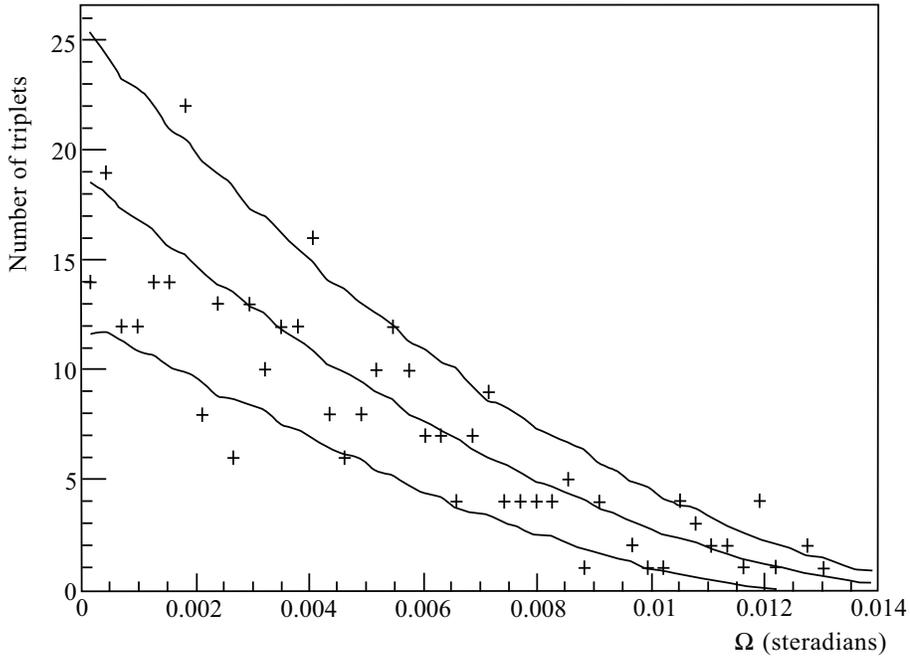}}
\caption[The distribution of solid angles, $\Omega$, of the spherical triangles.]{The distribution of solid angles, $\Omega$, of the spherical triangles. The points (+) represent the spherical triangles found in HiRes stereo data set. The lines represent the mean and the $\pm~1\sigma$ limits for the simulated data sets. The opening angles between each pair of points within the three points are required to be $< 10^{\circ}$.}
\label{cut10_spec}
\end{center}
\end{figure}

Figure~\ref{cut10_spec} shows the frequency with which groupings of events appear with a given spherical solid angle. The data are shown as (+) points while the solid lines mark the mean and the $\pm~1 \sigma$ limits from 1000 simulated data sets. The data lies well within the isotropic MC expectations, with no excess is seen at small solid angles. Note that the number of points (the data points) that lie outside the $\pm~1 \sigma$ are significantly less than $32 \%$ of the total number of points. This is due to the bin-to-bin correlation which we have studied and reported in~\cite{thesis}. The chance probability for the data to have at least as many triplets in the smallest angle bin (smallest $\Omega$) is calculated by counting the number of times the simulated data sets yield a stronger signal than the data. The chance probability of the signal to be greater than 14 (the value of the first bin from the HiRes data) is found to be equal to 0.75. Hence, no significant signal is found for small scale clustering of the HiRes stereo data set using the three-point correlation method.

\section{Three Point Correlation With BL Lac Objects}

As an alternate approach to find a possible correlation between the HiRes stereo data with the BL Lac objects, we used a modified three-point correlation method. A primary motivation for performing a multi-point correlation search with BL Lacs was that the two-point correlation search reported in~\cite{chad} was consistent with the experimental angular resolution of the HiRes detector without any magnetic smearing. A possible interpretation of this result was that the events correlated to the BL Lac objects are from neutral primaries, which is inconsistent with HiRes's report of proton-dominance of UHECR composition at the highest energies~\cite{greg}. In the three- and four-point correlation studies we test the more pedestrian hypothesis that the two-point correlation seen might simply be part of a more smeared out multi-event correlation. 

\begin{figure}[!h]
\begin{center}
{\includegraphics[width=12.0cm]{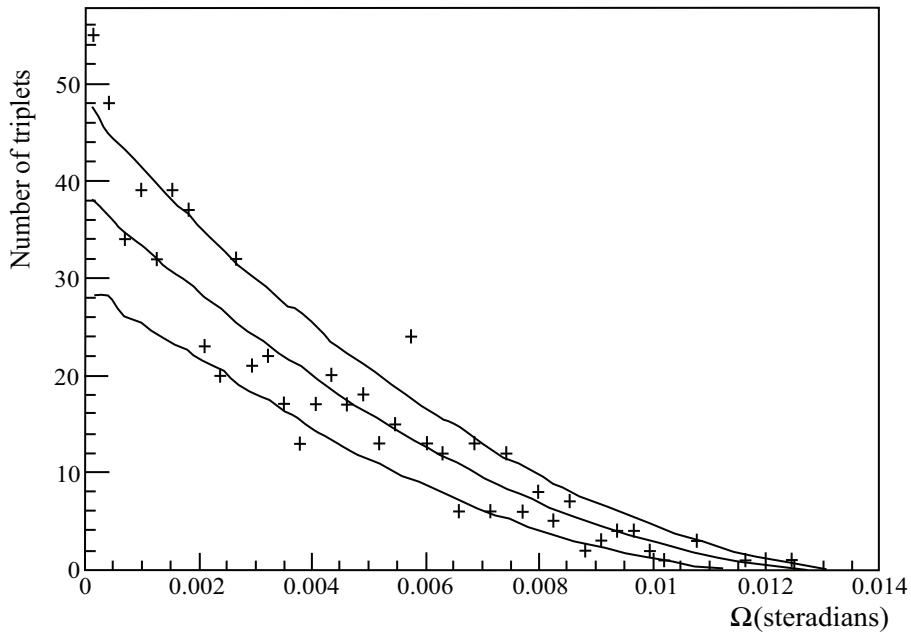}}
\caption[The distribution of areas of the spherical triangles.]{The distribution of areas for the spherical triangles found in using BL Lacs as the third point. The points (+) represent the spherical triangles using the HiRes stereo data set with the BL Lacs, where two points are from the data set and the third point is from the BL Lac location. The lines represent the mean and the $\pm~1\sigma$ for the simulated data sets with the same study. The opening angles between each two points of the three points are required to be $< 10^{\circ}$.}
\label{cut10_blhp_spec}
\end{center}
\end{figure}

In this study we calculated the solid angle of the spherical triangles contained between two event points from the data and one point from the BL Lac set.  As in the case of the self three-point correlation study presented above, we restricted our search to those events with angular separation less than $10^{\circ}$. Figure~\ref{cut10_blhp_spec} shows the distribution of solid angles from the spherical triangles found in the real data set shown as points (+) to be compared to the mean and $\pm~1 \sigma$ of 1000 simulated data sets shown as solid lines. As would be expected from the observation of a two-point correlation with BL Lacs. In the plot, we do see an apparent excess at small solid angles. Here the chance probability for the value of the real data to be greater than 55 (the value of the first bin) is found to be equal to 5.4\% ($\sim2\sigma$).

It is important to note that the significance of the three-point correlation is much weaker than that found in~\cite{chad} for the two-point correlation, which is consistent with what is expected from a random third point association with the existing correlated pairs. Therefore, no new information was found on BL Lac correlation with this study. 

\section{Cross Correlation Between the Centroid of the Triplets With Small Solid Angles and BL Lac Objects}

\begin{figure}[t]
\begin{center}
{\includegraphics[width=8.0cm]{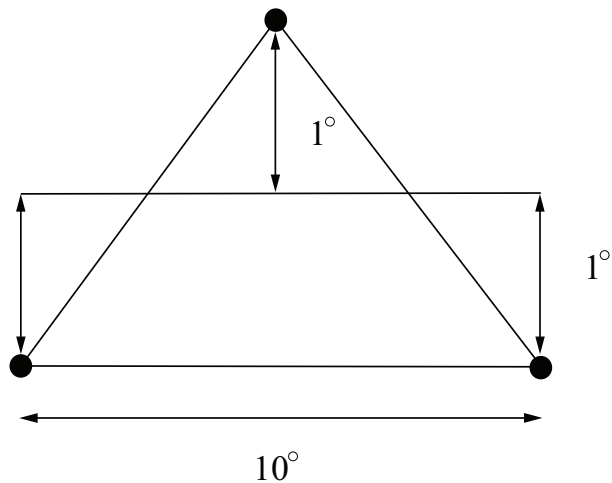}}
\caption{A diagram of a three points with a separation angle between the furthest two points of $10^{\circ}$. The middle point of the triangle is displaced one degree up and the end two points are displaced one degree (consistent with angular resolution smearing) down from where the three points would be placed on the same arc.}
\label{tmp1}
\end{center}
\end{figure}

In addition to the test in the previous section. We searched for cross-correlation  between the small solid angle clusters found in our data and the BL Lac objects. The cross-correlation is tested for triplets of solid angle $\Omega < 0.004$ sr. This cut corresponds to the selection of triplets that lie in an arc of up to $10^{\circ}$ and allows for a deflection of $1^{\circ}$ for the case where the middle event is centered between the two end events (consistent with the angular resolution of HiRes). Figure~\ref{tmp1} illustrates such a triplet. A total of 207 triplets were selected for study by this criterion.  The comparison was performed for the same set of BL Lacs used above.

\clearpage

\begin{figure}[!hpt]
\begin{center}
{\includegraphics[width=12.0cm]{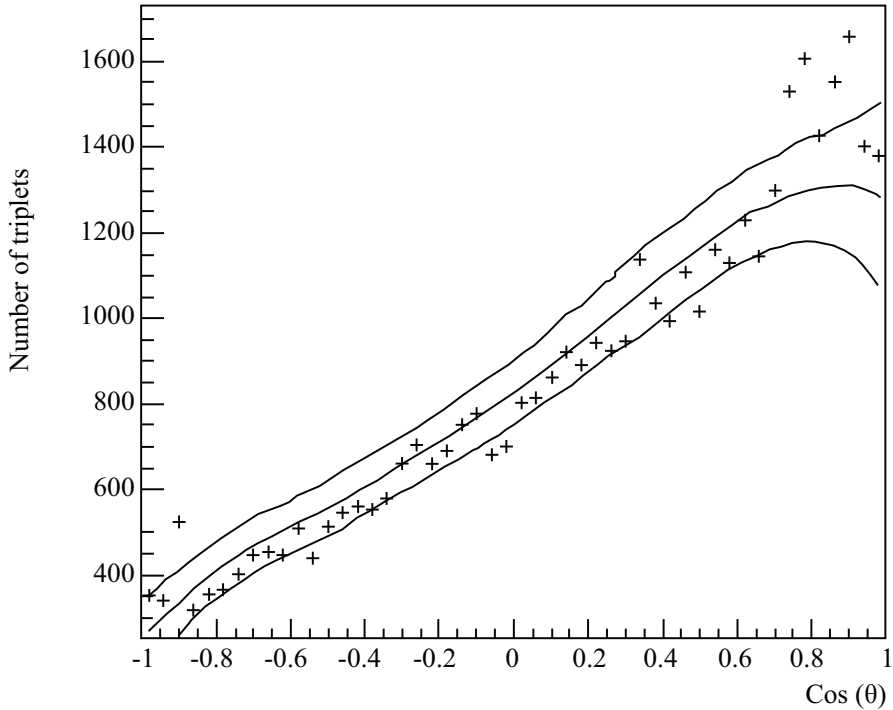}}
\caption{The frequency distribution of $cos(\theta)$, where $\theta$ is the opening angle between the middle point of the spherical triangles and the BL Lac. The spherical triangles have area, $\Omega < 0.004$ sr and the BL Lacs have magnitude $< 18$. The points (+) represent the real data set. The lines represent the mean and $\pm 1 \sigma$ limits calculated from 1000 isotropic MC sets.  }
\label{tmp4}
\end{center}
\end{figure}

Figure~\ref{tmp4} shows the histogram of $cos(\theta)$, where $\theta$ is the opening angle between the centroid point of a spherical triangles with solid angle, $\Omega < 0.004$ sr and the BL Lac set. The points (+) represent the data set and the solid lines show the mean and $\pm 1\sigma$ bounds from 1000 isotropically generated MC sets. As a figure of merit, we used the excess significance of the last bin where $cos(\theta)$ is closest to 1 ($\theta$ of $< 10^{o}$). Looking at the 1000 isotropic simulated data sets, we found the chance probability to be 0.3 of having an excess at least this great. 

Note that a few bins in this histogram display an excess of approximately $2\sigma$ in their occupancy level. However, one must remember that the values of the bins in this distribution are correlated. Therefore, each bin can not be treated independently. Consequently, the values of these bins can not be combined. Any excess seen is therefore no more than $2\sigma$  and consistent with the three-point correlation in the previous section. For this study we made the $\it{a~priori}$ choice to use the significance of the last bin and that is what we consistently use. Consequently, we conclude that there is no significant excess between the small solid angle triplets with the BL Lac objects.

\section{Four Point Correlation} 
\begin{figure}[!hpt]
\begin{center}
{\includegraphics[width=12.0cm]{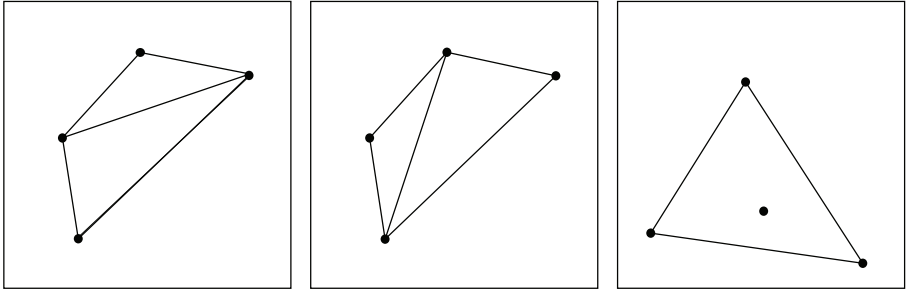}}
\caption[Divide.]{The first two plots describe two different divisions of a spherical quadrilateral into two spherical triangles. The same overall solid-angle is obtained from the sum of the two triangles, regardless of which division is used. The last plot shows the case where one of the points of quadruplet lies inside a spherical triangle bounded by the other three. In such case the solid-angle  is obtained from the spherical triangle containing the fourth point.}
\label{divide} 
\end{center}
\end{figure}

In the previous correlation study we looked for three-point correlation in the HiRes stereo data set using their spherical triangle area. In this section we will attempt to extend the technique to four-point correlations in an attempt to further investigate potential clustering in the HiRes stereo data set. We do this by calculating the solid-angle bounded by each quadruple set of events in the data.  The enclosed solid angle  is calculated by dividing the quadrilateral into two spherical triangles. The actual division can be done in two different ways, as illustrated in Figure~\ref{divide}. The total solid angle  is simply the sum of these two spherical triangles. Note that the value of the total solid angle is the same for either of the two divisions. In addition, the events can be configured such that a point of the quadruple is inside a spherical triangle formed by the other three points as shown in Figure~\ref{divide}. In such a case, the solid angle is that of the spherical triangle bounded by the outer three points. The resulting histogram of the solid angle areas is then compared to simulated data sets to estimate the significance of any excess.  

As in the case of the three-point autocorrelation study, we compare the distribution of the solid angles from the real data to that found in the simulated isotropic MC sets. The only difference is that we are using quadruplets rather than triplets. The solid angle of the quadruplets is calculated as the area of the exterior spherical triangles~\cite{thesis}. As in the case of the three point study, we have made a cut on the angular separation between pairs of points in the quadruples to be less than $10^{\circ}$.

\begin{figure}[!hpt]
\begin{center}
{\includegraphics[width=12.0cm]{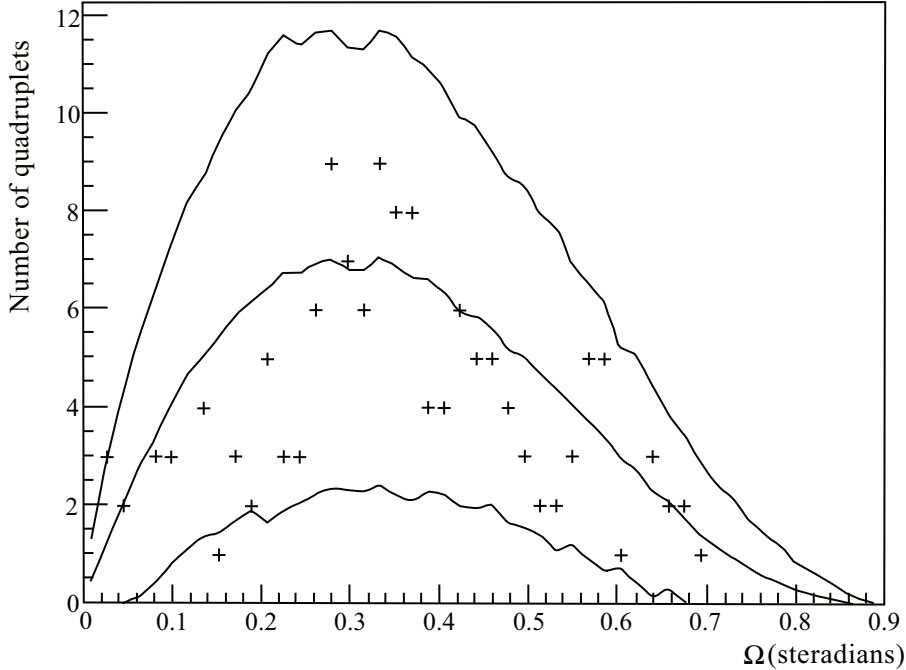}}
\caption[The histogram of the solid angle.]{The histogram of the solid angle between each of four points with the angular separation cut. The points (+) represents the real data sets, while the solid lines represent the mean and the $\pm~1\sigma$ of a 1000 simulated data set.}
\label{sp4_tot} 
\end{center}
\end{figure}

Figure~\ref{sp4_tot} shows the occupancy level vs. the solid angle area of quadruples with the $10^{\circ}$ angular separation cut. As previously, the actual HiRes data set is shown by the data points. The mean and $\pm~1\sigma$ intervals are shown by the solid lines. The figure clearly shows no obvious excess in the lowest bins. Again, the chance probability of the first bin from 1000 MC sets area (corresponding to the smallest solid angle) is again calculated by dividing the number of simulated data sets that yielded a stronger signal than that of the real data set by the total number of simulated data set. In this case, the value was found to be $\sim 0.70$. Hence, clearly no significant four-point clustering is found in the data. 

When looking at the distribution of data points in Figure~\ref{sp4_tot}, they are found to be systematically lower when compared with the mean obtained from the simulated sets. This deficit suggests that our understanding of the detector aperture is not good enough to reproduce the data beyond a four-point correlation search. In addition, the number of clusters of small angle four-point cluster is barely significant for statistical analysis. Hence, we decided in this study to stop at four-point correlation and did not proceed to five (or more) point auto correlation. 

\section{Conclusion}
In this paper we described a search for small scale clustering in the HiRes stereo data set using three point correlation. The search is applied by histogramming the area subtended by spherical triangle defined by triplets in the HiRes stereo data set. We required that every pair in a triplet lie within $10^{\circ}$ of angular separation. The result from the histogram is then compared to that found in the simulated data sets. The simulated data sets were generated to have the same exposure as that of the real observed data. No excess of small scale clustering in the HiRes stereo data set was found.

We also looked for correlation between the HiRes stereo data set and the BL Lac objects using the three point correlation technique. The strategy used in this study is looking for three points where two of the three points are from the data set and the third one is a BL Lac location, each of which has a maximum angular separation of $10^{\circ}$. The significance found is approximately $2 \sigma$ and is weaker then the the two-point correlation study in~\cite{chad}. We conclude that no new information on BL Lac correlation was obtained from these studies. In addition, we looked for cross correlation between the HiRes stereo data triplets with $\Omega < 0.004$ sr and the BL Lac set. No significant correlation was found between the triplets and the BL Lac objects. 

Finally, we tested for four point correlation in the HiRes stereo data set and found no statistically significance excess. We conclude that the four point auto correlation study is pushing both statistical borderline and systematic limits, of the reliability of the simulation. For this reason, we decided not to extend the study to four point BL Lac correlation or to five point auto correlation.

\section{Acknowledgments}
This work is supported by the National Science Foundation under contracts NSF-PHY-9321949, NSF-PHY-9322298, NSF-PHY-9974537, NSF-PHY-0071069, NSF-PHY-0098826, NSF-PHY-0140688,  NSF-PHY-0245328, NSF-PHY-0307098,  and NSF-PHY-0305516, as well as by Department of Energy grant FG03-92ER40732. We gratefully acknowledge the contribution from the technical staffs of our home institutions. We gratefully acknowledge the contributions from the University of Utah Center for High Performance computing. The cooperation of Colonels E. Fisher, G. Harter, and G. Olsen, the US Army and the Dugway Proving Ground staff is appreciated.   



\begin{thebibliography}{9}
   

\bibitem{chad}
R.U.~Abbasi \textit{et al.}, Astrophys. J.636, p.680, 2006.

\bibitem{agasa}
Takeda \textit{et al.}, Astrophys. J.522, p.225, 1999.

\bibitem{auger}
The Pierre Auger Collaboration, Science V.318, p.938, 2007.

\bibitem{hough}
R.U.~Abbasi \textit{et al.}, Astrophys.  Volume 28, Issue 4-5, p385, 2007.


\bibitem{ch6-3}
R.C.~Hartman \textit{et al.}, Astrophys.~J.Suppl. 123, \textbf{79} (1999).

\bibitem{ch6-4}
P.G.~Tinyakov \& I.I.~Tkachev, JETP Lett., 74, 445, 2001.

\bibitem{ch6-5}
P.G.~Tinyakov \& I.I.~Tkachev, Astropart. Phys., 18, 165, 2002.

\bibitem{ch6-6}
D.S.~Gorbunov, P.G.~Tinyakov, I.I.~Tkachev, \& S.V.~Troitsky, Astrophys. J.577, L\textbf{93}, 2002.

\bibitem{ch6-7}
D.S.~Gorbunov, P.G.~Tinyakov, I.I.~Tkachev, \& S.V.~Troitsky, JETP Lett., 80, 145, 2004. 


\bibitem{data}
R.U.~Abbasi \textit{et al.}, Astrophys. J.610, L\textbf{73}, 2004.

\bibitem{ch8}
http://math.rice.edu/~pcmi/sphere/gos4.html

\bibitem{ch8-1}
W.M.~Smart, \textit{Spherical Astronomy}, Cambridge University Press, 1960.


\bibitem{thesis}
R.U.~Abbasi, Ph.D. thesis, University of Utah, 2007.

\bibitem{greg}
R.U.~Abbasi \textit{et al.}, Astrophys. J.622, p.910, 2005.


\end{thebibliography}
\end{document}